# A polymer brush theory for quantitative prediction of maximum height change between dry and wet states

Jiawei Yang*

David H Koch Institute for Integrative Cancer Research, Massachusetts Institute of Technology, Cambridge, MA 02138, USA
Department of Anesthesiology, Boston Children's Hospital, Boston, MA 02115, USA.

*To whom correspondence should be addressed. Email: jwyang@mit.edu

**Abstract**

Polymer brushes can grow on almost any solid surface, and by design, exhibit diverse properties and functionalities, thus they have been widely used in many emerging applications in engineering, energy, and medicine. In particular, some applications such as actuation, molecule release, and friction switch require the polymer brushes to change their heights between dry and wet states, and maximizing such height change is critical for the optimal performance of these applications. While scaling laws have long been proposed to qualitatively determine brush heights, a theory that can quantitatively predict brush heights and conditions for maximizing brush height change is still lacking yet is valuable for the practical design of polymer brushes. Here, we take a thermodynamic approach to formulate a polymer brush theory to calculate brush heights at various conditions of graft area, degree of polymerization (DP), and solvent qualities. Our model consists of two parts—the freely-jointed chain model to describe the elasticity of brushes and the Flory-Rehner model to describe the mixing of brushes and solvents. The calculated brush heights at both dry and wet states fairly agree with the experimental data from the literature. The calculated brush heights are further used to determine the conditions for the maximum brush height change. Our theory can guide the design of polymer brushes for optimal functional performance in various applications and also can couple with other models to describe more complex behaviors of polymer brushes.

Polymer brushes are a layer of polymer chains with one end anchored on the surface of a substrate and the other end free. Polymer brushes can grow on most surfaces of solids of different materials and geometries, with thicknesses ranging from a few nanometers to a few micrometers. Polymer brushes can be tailored with desired chemical, physical, and biological properties to enable surface functions that are distinct from the substrate. These unique features thrust rapid and intensive research to leverage polymer brushes to create a broad range of emerging applications in engineering, energy, and medicine [1-4]. Examples include adhesion, friction, and lubrication controls [5-7], sensors [4,8] and actuators [9,10], high-performance batteries [11], antifouling surfaces [12,13], and immunomodulation coatings [14].

While polymer chemistry mostly determines the brush functionality, polymer physics determines the brush conformations [15,16]. The combination of functionality and conformation gives rise to their overall performance. In particular, the change in brush heights as a direct consequence of the change in brush conformations can strongly influence the performance of applications. Indeed, many applications require a large change of brush height between dry and wet states to realize optimal performance. For example, in a binary polymer brush system, two types of brushes should exhibit an opposite maximum height change between the swollen and collapsed states to switch the friction behavior [17]; in a diblock copolymer brush system, to regulate the molecule release, the top block of brushes should extend most to allow the release while completely collapse to block the release [18]; in a polymer brush actuation system, dry brushes coated on a cantilever should swell most to achieve a maximum deflection [19]. Therefore, a design guideline for polymer brushes that can maximize their height change between wet and dry states is demanded.

Various theories of polymer brushes have been proposed and used for decades [15,20-22]. Alexander [23] and de Gennes [24] among the pioneer researchers first derived a set of scaling laws to determine brush conformations on brush graft density, brush length, and brush-brush and brush-solvent interactions. Since then, great efforts have been devoted to gaining a deeper understanding of polymer brushes by developing new theories and simulations, such as the "blob concept" [25], Monte Carlo simulations [26], and Brownian dynamic simulations [27], and investigating various aspects of brush physics, such as the effect of charged brushes [28], the effect of solvent compositions (e.g., mixed solvents, semi-dilute solvent, and polymer melt) [20,29], the monomer distribution along the brush, and the brush-end distribution [30,31]. However, a theory that can quantitatively calculate brush heights is still lacking.

Here, we take a thermodynamic approach to formulate a polymer brush theory to calculate the brush heights under varied graft areas, degree of polymerization (DP), and solvent qualities, and further determine the brush height change between dry and wet states. An individual polymer brush is confined in an occupied volume defined by the graft area and the brush height. We use the freely-jointed chain model to describe the elasticity of a polymer brush and the Flory-Rehner model to describe the mixing of brushes and solvents. Brush heights are calculated when brushes are in equilibrium with solvents. We show that brush heights predicted by our model fairly agree with experimental data from the literature. The maximum height change between dry and wet states is found at an intermediate graft area, weakly dependent on the solvent quality, and increases monotonically with DP.

Consider a layer of polymer brushes anchored on a substrate, of height $H$ and separated by a spacing $s$. Each polymer brush consists of repeating units—the monomers—that are joined by covalent bonds. Let the length of a monomer be $a$ and the number of the repeating unit (i.e., DP) be $n_p$, then the brush contour length is $n_p a$. When $H \ll s$, brushes do not interact with each other, independent of $s$ (**Figure 1a, (i)**), and $H$ can be estimated by the Flory radii [32]; when $H \sim s$, brushes make contact and become entangled and overlapped (**Figure 1a, (ii)**). Due to the steric confinement from the neighbor brushes (also known as the excluded volume repulsion), brushes favor stretching themselves to a bigger height to reduce the interaction rather than forming further entanglement; when $s \to a$ and $H \gg s$, brushes stretch almost to the full extension $H \to n_p a$ (**Figure 1a, (iii)**). The evolution of brush height with brush spacing is summarized in **Figure 1b**. From dry brushes to wet brushes, the increase of brush height $\Delta H$ depends on $s$. Dilute brushes can stretch themselves to the height of their Flory radii, and $\Delta H$ is small; dense brushes are already highly stretched so that $\Delta H$ is also small. Therefore, the maximum $\Delta H$ should be found at an intermediate $s$ (**Figure 1c**).

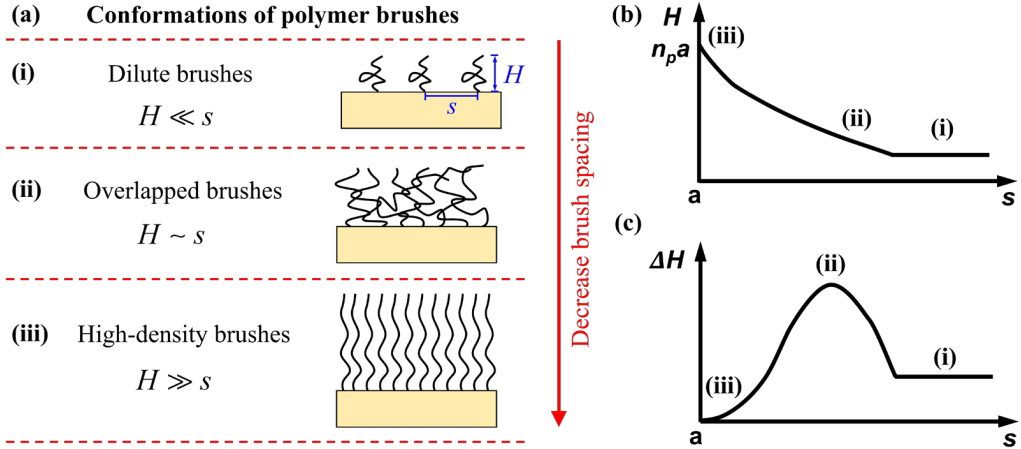

**Figure 1. Polymer brush height evolves with conformations. (a)** Brush conformations depend on brush spacing. Schematic illustration of **(b)** $H$ and **(c)** $\Delta H$ evolving with $s$. Three brush conformations are marked on the curves. The maximum $\Delta H$ should be found at an intermediate $s$.

To formulate the thermodynamic model, we make the following assumptions (i) all brushes are of the same height and spacing, (ii) brushes are mainly stretched in the normal direction and the entanglement with the neighbor brushes is ignored, and (iii) the distributions of monomers and solvent molecules along the brush are uniform so that their interactions are same. The layer of polymer brushes can be equally divided into individual brushes and each brush is confined in an occupied volume of $H\sigma$ ($\sigma = s^2$, the graft area per brush) (**Figure 2a**). The volume conservation requires

$$n_s a^3 + n_p a^3 = H\sigma \tag{1}$$

where $n_s$ is the number of solvent molecules. The size of a solvent molecule is assumed same as that of a monomer.

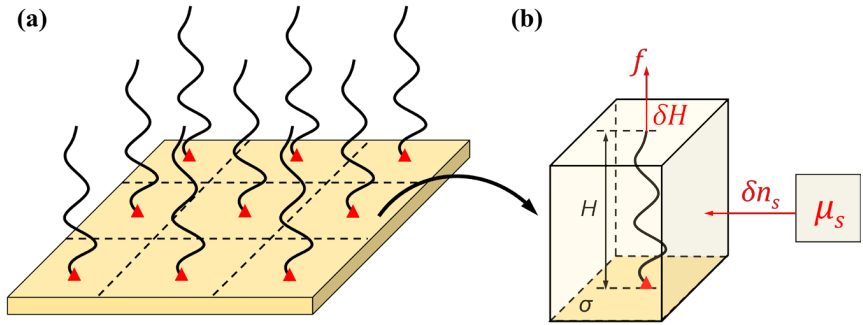

**Figure 2. Polymer brush model. (a)** A layer of polymer brushes can be equally divided into individual brushes with the same occupied volume of area $\sigma$ and height $H$. **(b)** In each occupied volume, the brush is subject to a mechanical load $f$ and is in contact with a solvent of a constant chemical potential $\mu_s$.

The brush is subject to a mechanical load $f$ and is in contact with a solvent of a constant chemical potential $\mu_s$ (**Figure 2a**). The brush, the mechanical load, and the chemical load form a thermodynamic system. In this thermodynamic system, the constitutive law of brushes can be described by free energy functions, which involve two molecular processes: the stretch of the brush and the mixing of the brush with solvent molecules. The free energy function can be written as

$$W(H, n_s) = W_{str}(H) + W_{mix}(n_s) + \Pi(n_s a^3 + n_p a^3 - H\sigma) \tag{2}$$

where $W_{str}$ and $W_{mix}$ are the free energies of stretching and mixing. The last term enforces the volume conservation (1) and $\Pi$ is the Lagrange multiplier to be determined by the boundary condition. This free energy is a function of two variables, $H$ and $n_s$. Associated with small changes, $\delta H$ and $\delta n_s$, the free energy changes by

$$\delta W(H, n_s) = \left(\frac{\partial W_{str}(H)}{\partial H} - \Pi\sigma\right)\delta H + \left(\frac{\partial W_{mix}(n_s)}{\partial n_s} + \Pi a^3\right)\delta n_s \tag{3}$$

The free energy change is caused by the work done by the mechanical load and the chemical load, namely,

$$\delta W(H, n_s) = f\delta H + \mu_s \delta n_s \tag{4}$$

Comparing Eq. (3) and Eq. (4) yields

$$f = \frac{\partial W(H, n_s)}{\partial H} \tag{5a}$$

$$\mu_s = \frac{\partial W(H, n_s)}{\partial n_s} \tag{5b}$$

We adopt the freely-jointed chain model to write the free energy of stretching

$$W_{str}(H) = n_p kT \left(\frac{H}{n_p a}\beta + \log\frac{\beta}{\sinh\beta}\right) \tag{6}$$

where $k$ is the Boltzmann constant and $T$ is temperature. $\beta = L^{-1}(H/n_p a)$ is a measure of force, and $L^{-1}$ is the inverse Langevin function, defined by $L(x) = \coth(x) - 1/x$. $H/n_p a$ is a measure of stretch, confined between 0 and 1. For a single brush, the brush is in an unperturbed state when $H = a\sqrt{n_p}$, and $H/n_p a = 1/\sqrt{n_p}$; when $H/n_p a > 1/\sqrt{n_p}$, the brush stretches; as $H/n_p a \to 1$, the brush straightens to a full extension and $\beta$ diverges; when $H/n_p a < 1/\sqrt{n_p}$, the brush collapses. In particular, for small deformation, i.e., $\beta \ll 1$, $W_{str}(H) = kT(3H^2/2n_p a^2)$, which has long been implemented in theoretical modeling of polymer brushes [6,15,29]. Here our free energy function does not have the limitation of small deformation and can be applied to arbitrary large deformation.

Polymer brushes do not have translational entropy as they are linked through a giant substrate. Therefore, we adopt the Flory-Rehner model to describe the free energy of mixing

$$W_{mix}(n_s) = kT(n_s \log\phi_s + \chi n_s \phi_p) \tag{7}$$

where the first term is the entropy of mixing and the second term is the enthalpy of mixing. $\chi$ is the Flory constant which describes the affinity between brush monomers and solvent molecules. $\phi_s$ and $\phi_p$ are the volume fractions of the solvent molecules and the polymer brush in the occupied volume, namely,

$$\phi_s = \frac{n_s}{n_s + n_p}, \quad \phi_p = 1 - \phi_s = \frac{n_p}{n_s + n_p} \tag{8}$$

Recall Eq. (1), $\phi_p$ can also be expressed as

$$\phi_p = \frac{n_p}{\bar{H}\bar{\sigma}} \quad (9)$$

here $\bar{H} = H/a$, $\bar{\sigma} = \sigma/a^2$, dimensionless forms of $H$ and $\sigma$. The force at the end of the brush is calculated using Eq. (5a)

$$f = \frac{kT}{a} L^{-1}\left(\frac{H}{n_p a}\right) - \Pi\sigma \quad (10)$$

here $\Pi$ is the osmotic pressure. The chemical potential of solvent molecules is calculated using Eq. (5b)

$$\mu_s = kT\left[\log(1-\phi_p) + \phi_p + \chi\phi_p^2\right] + \Pi a^3 \quad (11)$$

At the equilibrium state, Eq. (10) and Eq. (11) both vanish. By eliminating $\Pi$, we obtain a governing equation

$$\log(1-\phi_p) + \phi_p + \chi\phi_p^2 + \frac{1}{\bar{\sigma}} L^{-1}\left(\frac{1}{\phi_p \bar{\sigma}}\right) = 0 \quad (12)$$

Given $\bar{\sigma}$ and $\chi$, Eq. (12) can be numerically solved to calculate $\phi_p$. By substituting $\phi_p$ into Eq. (9), $\bar{H}$ can be determined.

We first calculate the polymer brush ($n_p$ = 1000) volume fractions and heights in a good solvent ($\chi$ = 0.45) at varied graft areas across multiple orders of magnitude (**Figure 3a** and **b**). When $\bar{\sigma} \to 1$, the graft area is about the size of a monomer. Brushes stretch almost to a full extension, reaching the high-density limit, therefore $\phi_p \to 1$ and $\bar{H} \to n_p$. As $\bar{\sigma}$ increases, brushes are less crowded but maintain overlapped. A bigger graft area allows more solvent molecules to mix with the brushes, leading to a smaller $\phi_p$ and a larger $\bar{H}$. $\bar{H}$ scales with $\bar{\sigma}$ of power -0.4. As $\bar{\sigma}$ further increases beyond a critical condition $\bar{\sigma} = \bar{H}^2$, where the brush height equals the brush spacing, brushes enter the dilute region and become non-contact. $\bar{H}$ is constant, same as that when $\bar{\sigma} = \bar{H}^2$, independent of $\bar{\sigma}$. From Eq. (9), $\phi_p$ decreases with $\bar{\sigma}$ through the relation $\phi_p \sim 1/\bar{\sigma}$.

We then investigate the effect of solvent quality. A better solvent enhances the brush-solvent mix, resulting in a smaller $\phi_p$ and a bigger $\bar{H}$ (**Figure 3c** and **d**). When $\bar{\sigma} \to 1$, all curves converge to the same values, $\phi_p \to 1$ and $\bar{H} \to n_p$, independent of $\chi$, as the extremely small graft area cannot accommodate solvent molecules. In the overlapped brush region, $\phi_p$ scales with $\bar{\sigma}$ of power -2/3, -1/2, and 0 in a good solvent ($\chi$ = -0.45), a $\theta$-solvent ($\chi$ = 0.5), and a poor solvent ($\chi$ = 1) respectively. Recall Eq. (9), $\bar{H}$ scales with $\bar{\sigma}$ of power -1/3, -1/2, and -1, which are consistent with the results from the scaling analysis [15,20,33]. In the dilute brush region, the brushes do not interact for all solvents. $\bar{H}$ is constant and $\phi_p \sim 1/\bar{\sigma}$. These scaling relations fairly agree with experimental observations in poly(acrylamide) (PAAM) brushes [34], poly(methyl methacrylate) (PMMA) brushes [35], poly(ethylene glycol) (PEG) brushes [36], and poly(styrene) (PS) brushes [37].

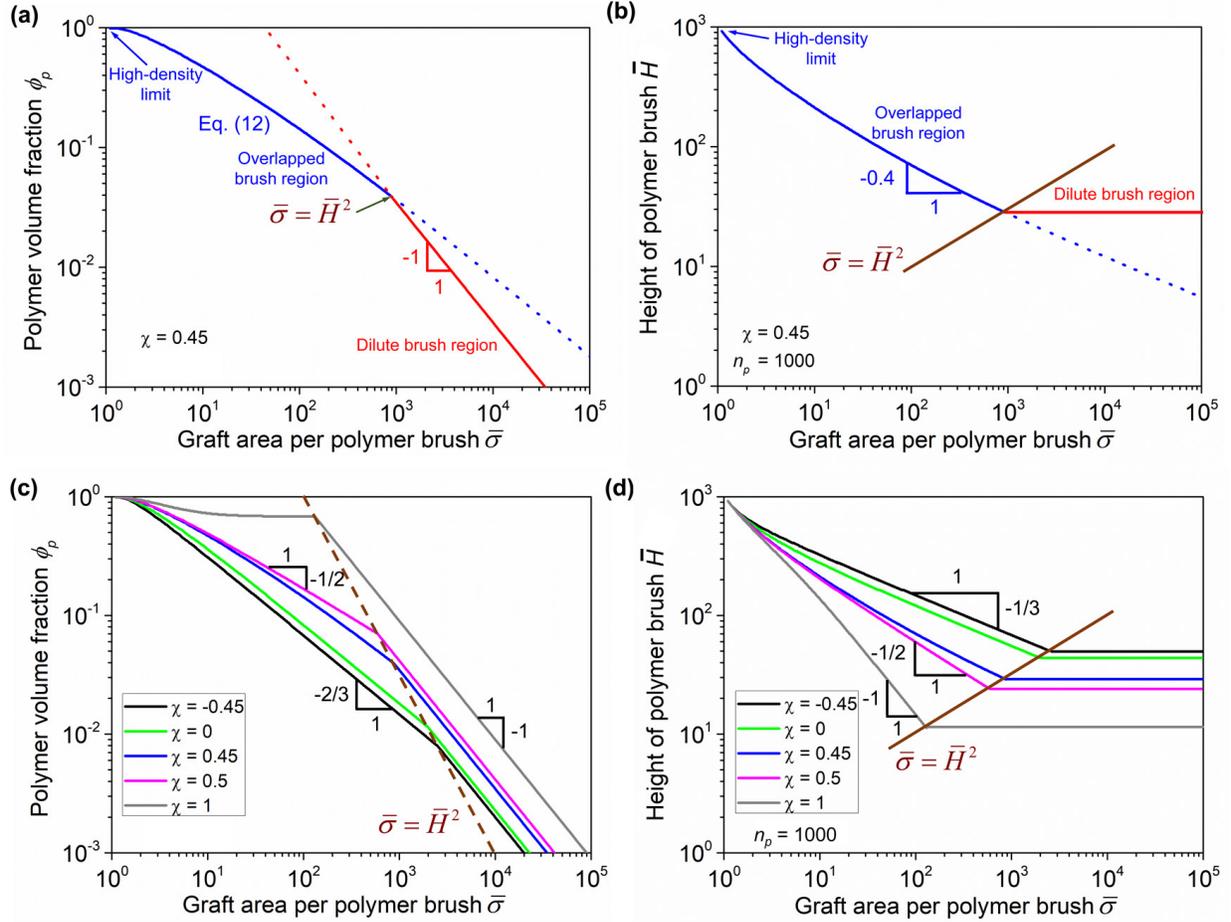

**Figure 3. Polymer brush volume fractions and heights as a function of graft areas in different solvent qualities.** (a) $\phi_p$ and (b) $\bar{H}$ as a function of $\bar{\sigma}$ when $n_p = 1,000$ and $\chi = 0.45$, and (c-d) in various $\chi$.

We further study the effect of DP. We fix $\bar{\sigma} = 100$, and thus $\bar{H} = 10$ is the boundary between the overlapped brushes and dilute brushes. $\bar{H}$ is same as that of Flory radius when $\bar{H} < 10$ and linearly increases with $n_p$ when $\bar{H} > 10$ (**Figure 4a**). A better solvent leads to a bigger brush height at the same $n_p$. Brushes of a smaller $n_p$ in a good solvent can have the same height as those of a bigger $n_p$ in a poor solvent. We also vary $\bar{\sigma}$ from 1 to 100000, $\bar{H}$ scales with $n_p$ of power 1 at $\bar{\sigma} = 1$ and 0.55 at $\bar{\sigma} = 100000$ respectively (**Figure 4b**), that is because the brush spacing is either too small or too big compared to the brush height in the given range of $n_p$ and the brushes always stay in the same state. In between the two limits, transition $n_p$ exists, below which brushes are dilute and above which brushes are overlapped. Brushes of a smaller $n_p$ with a smaller graft area can have the same height as those of a bigger $n_p$ with a bigger graft area.

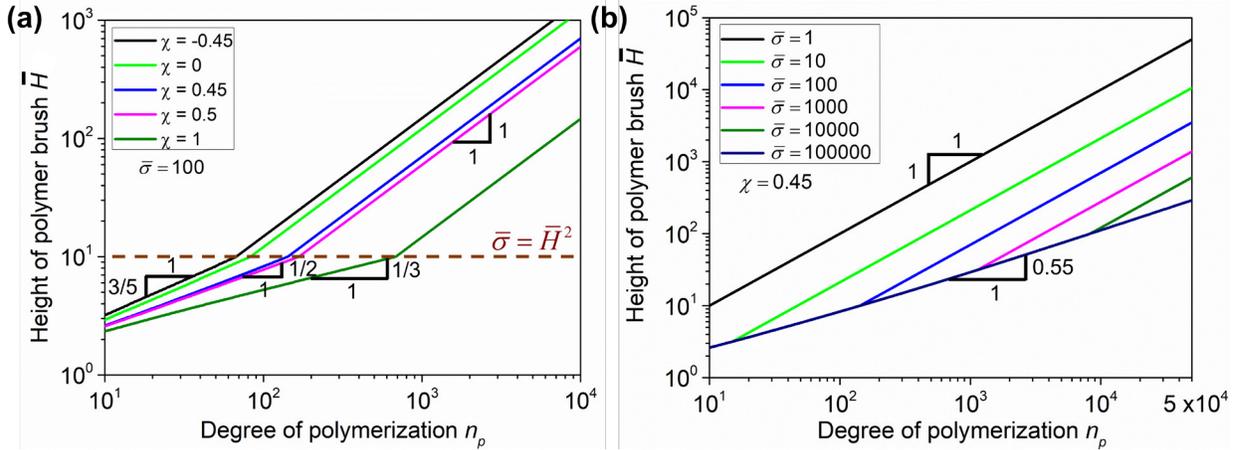

**Figure 4. Polymer brush height as a function of the degree of polymerization.** $\bar{H}$ as a function of $n_p$ **(a)** under various $\chi$ when $\bar{\sigma} = 100$ and **(b)** under various $\bar{\sigma}$ when $\chi = 0.45$.

We next determine the brush heights as a function of graft areas in the dry state. For overlapped and high-density brushes, we set $\phi_p = 1$, Eq. (9) becomes

$$\bar{H}_d = n_p / \bar{\sigma} \tag{13}$$

here $\bar{H}_d$ is the dry brush height. As the dry brushes enter the dilute region, $\bar{H}_d$ is same as that at $\bar{\sigma} = \bar{H}_d^2$. Thus, from Eq. (13), the dilute brush height is

$$\bar{H}_d = n_p^{1/3} \tag{14}$$

We validate our model with experimental data from the literature. Neutral polymer brushes are selected to satisfy our model to exclude strong brush-brush interactions. Polymer brushes selected include poly(acrylic acid) (PAA) [7], poly(ethylene oxide) (PEO) [38], poly(ethyl methacrylate) (PEMA) [39], poly(2,2,2-trifluoroethyl methacrylate) (PTFEMA) [39], poly(tert-butyl acrylate) (PtBA) [40], PMMA [35], PAAM [34], and PS [37] (**Figure 5**). We convert experimentally measured $\sigma$ to $\bar{\sigma}$ by $\bar{\sigma} = \sigma / a^2$. Here $a$ is estimated as $(M/A\rho)^{1/3}$, where $M$ is the molecular weight of the monomer, $\rho$ is the density of the monomer, and $A$ is the Avogadro number (6.022 × 10$^{23}$ mol$^{-1}$). For PAAM, PAA, and PEO, $a$ = 0.47 nm, 0.485 nm, and 0.4 nm. As a result, our model predicts well for the dry brush heights and fairly agrees with the wet brush heights, albeit some of $\chi$ values are unknown. For example, the heights of PMMA brushes in acetone (a good solvent), in the mixture of 60% acetone and 40% methanol (a $\theta$-solvent), and in methanol (a poor solvent) exactly follow the calculated curves with $\chi$ = 0, 0.5, and 1. While for other brushes such as PEMA and PtBA in acetone and dimethylformamide (DMF) (both are good solvents), the calculated brush heights are in the good solvent range of $\chi < 0.5$. In addition, PTFEMA in acetone, PS in toluene ($\chi$ = 0.37), and PAAM in water ($\chi$ = 0.495) exhibit higher brush heights than the calculations. It is possibly because (1) our model is too simplified to capture the behaviors of all types of brushes, and (2) these polymer brushes have a high polydispersity in experiments (e.g., polydispersity index = 1.7 for PAAM brushes) such that the brush heights themselves are non-uniform.

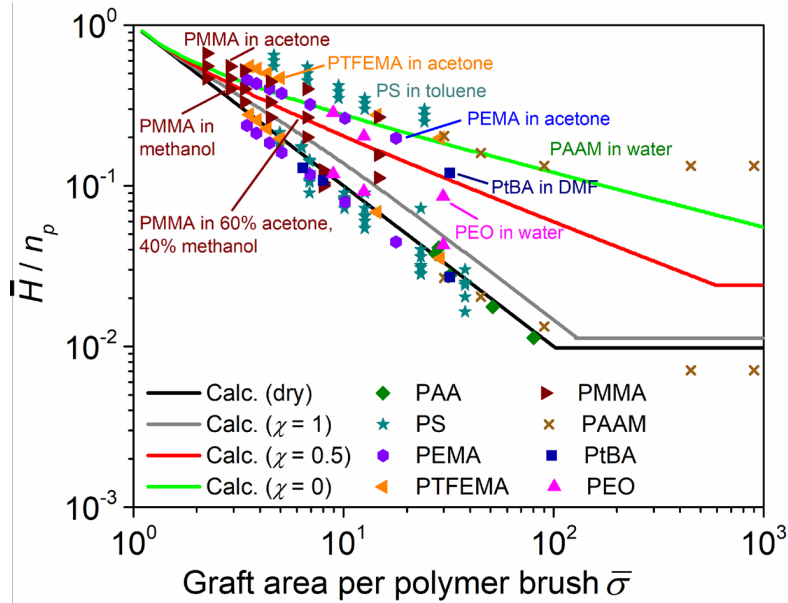

**Figure 5.** Comparison of calculated and experimentally measured polymer brush heights in dry and wet states.

We calculate the brush height change $\Delta \bar{H} = \bar{H} - \bar{H}_d$ between the dry state and the wet state at different graft areas. When dry brushes at a certain graft area are wet in a good solvent (e.g., $\chi = 0.45$), brush height increases, but the amount of increase varies, indicated by the gap between the two curves (**Figure 6a**). At a small $\bar{\sigma}$ ($\bar{\sigma} \to 1$, high-density brush), the increase is negligible, while at a big $\bar{\sigma}$ ($\bar{\sigma} > 1000$, dilute brush), the increase is constant but small, $\Delta \bar{H}/n_p = 0.019$. A maximum height increase is found at an intermediate $\bar{\sigma}$, where $\Delta \bar{H}/n_p = 0.116$ at $\bar{\sigma} = 7.2$, almost by an order of magnitude increase. The calculation is consistent with our hypothesis (**Figure 1c**). A better solvent quality leads to a bigger increase in brush height, but the corresponding $\bar{\sigma}$ is almost insensitive to solvent qualities (**Figure 6b**). For example, maximum $\Delta \bar{H}/n_p = 0.227$ and 0.109 are found at $\bar{\sigma} = 8.6$ and 6.4 for $\chi = -0.45$ and 0.5.

In addition, we calculate $\Delta \bar{H}$ at different $n_p$ under fixed $\bar{\sigma} = 100$. In a good solvent (e.g., $\chi = 0.45$), the brush height increases with $n_p$ for both dry and wet brushes and the gap between the two curves monotonically increases with $n_p$ (**Figure 6c**). Similarly, a better solvent quality leads to a bigger brush height increase. Maximum $\Delta \bar{H}$ increases with $n_p$ (**Figure 6d**). Such a tendency is general for arbitrary $\bar{\sigma}$.

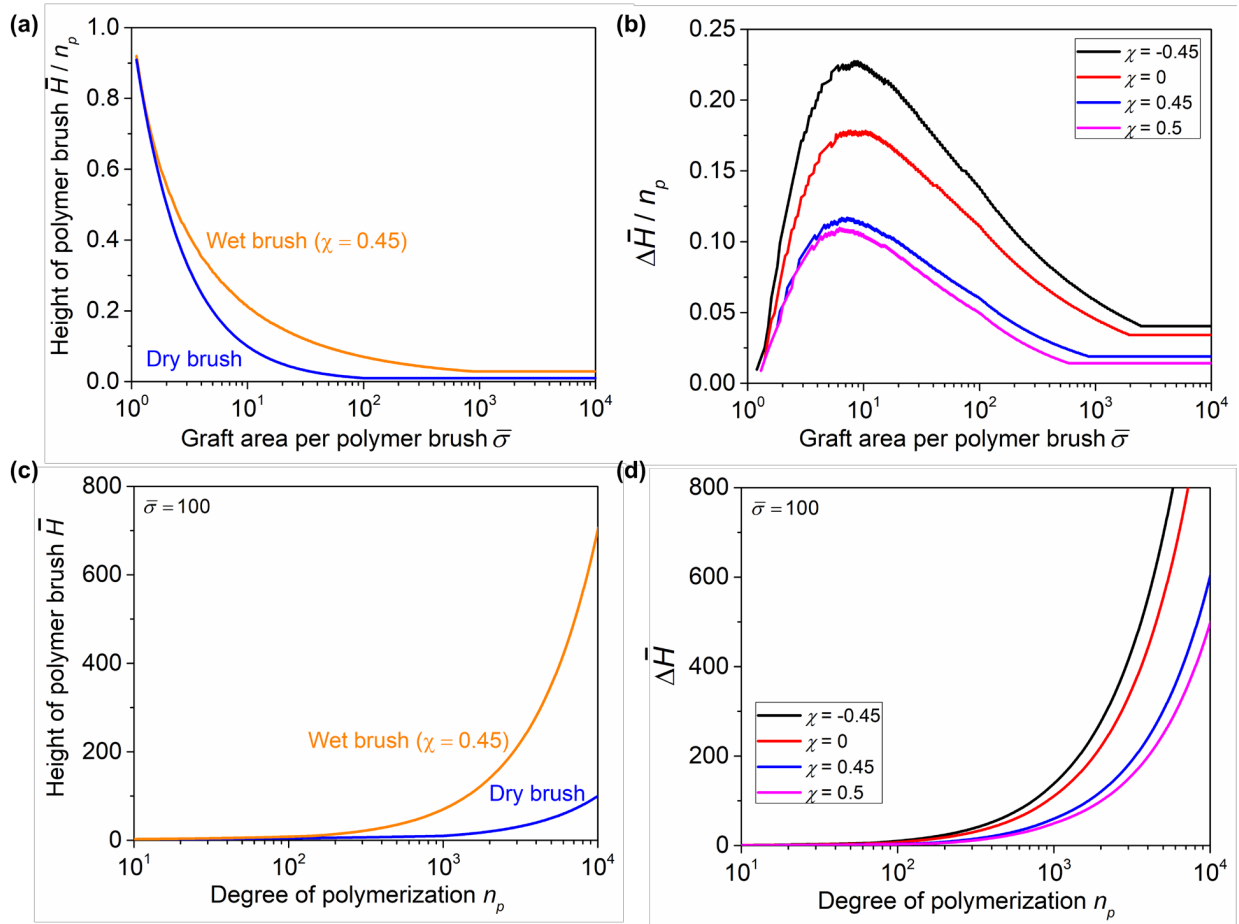

**Figure 6. Change of polymer brush height between dry and wet states. (a)** $\bar{H}/n_p$ for dry and wet brushes as a function of $\bar{\sigma}$. **(b)** $\Delta\bar{H}/n_p$ as a function of $\bar{\sigma}$ under various $\chi$. The maximum change in brush height is found at an intermediate $\bar{\sigma}$. **(c)** $\bar{H}$ for dry and wet brushes as a function of $n_p$. **(d)** $\Delta\bar{H}$ as a function of $n_p$ under various $\chi$. The maximum brush height increases with $n_p$.

The practical use of our polymer brush model may be limited by the assumptions, but such a simplified model does predict the brush heights at various conditions of graft areas, DP, and solvent qualities. Indeed, to better describe the polymer brush behaviors, a more elaborated model is needed, which should account for the monomer distribution and the monomer-solvent interaction along the brush, the brush-brush interactions, and the brush-brush entanglements. In addition, to model polymer brushes of strong interactions, detailed information about individual interactions should be added to the current model, such as electrostatic interaction for charged brushes and environmentally coupled interactions (e.g., temperature-responsive, pH-responsive) for responsive brushes. Developing such models is our future focus.

Our model does not limit to applications that require the maximum height change between dry and wet states, it can also be implemented in applications that benefit from the maximum height change. Indeed, many experiments have already reported that the optimal functional performance is observed at an intermediate brush thickness and graft density. For example, poly(sulfobetaine) (PSBMA) brushes resist protein absorption most at a moderate brush thickness of 62 nm [41]; poly(hydroxyethyl methacrylate) (PHEMA) brushes give the lowest marine bacterium attachment at a brush thickness of 20-40 nm, which corresponds to a maximum swelling of brushes (or maximum height increase) [42]; poly(dimethylsiloxane) (PDMS) brushes achieve the maximum adhesion energy at an intermediate graft density of 0.01, weakly dependent on the DP of the brush [43]. It is interesting to incorporate models of relevant physical processes in these applications to our model to predict their functional outcomes.

In summary, we formulate a polymer brush theory to quantitatively predict the brush height and the maximum brush height change between dry and wet states at various conditions of graft area, DP, and solvent qualities. We use the freely-jointed chain model and the Flory-Rehner model to describe polymer brushes equilibrated in solvents. The calculated brush heights fairly agree with the experimental data from the literature. The maximum brush height increase from the dry state to the wet state is found at an intermediate graft area, weakly dependent on the solvent quality, and increases monotonically with DP. This model can guide the design of polymer brushes for optimal performance in many practical applications and can be further used to couple with other models to describe more complex behaviors of polymer brushes.

**Acknowledgment**
This work was supported by the Leona M. and Harry B. Helmsley Charitable Trust Foundation Grant (2017PG-T1D027).